\documentclass[aps,twocolumn]{revtex4}
\begin{document}
\title{Extension of the $CPT$ Theorem to non-Hermitian Hamiltonians and Unstable States}
\author{Philip D. Mannheim}
\affiliation{Department of Physics, University of Connecticut, Storrs, CT 06269, USA.
email: philip.mannheim@uconn.edu}
\date{December 10, 2015}
\begin{abstract}
We extend the $CPT$ theorem to quantum field theories with non-Hermitian Hamiltonians and unstable states. Our derivation is a quite minimal one as it requires only the time-independent evolution of scalar products, invariance under complex Lorentz transformations, and a non-standard but nonetheless perfectly legitimate interpretation of charge conjugation as an antilinear operator. The first of these requirements does not force the Hamiltonian to be Hermitian. Rather, it forces its eigenvalues to either be real or to appear in complex conjugate pairs, forces the eigenvectors of such conjugate pairs  to be conjugates of each other, and forces the Hamiltonian to admit of an antilinear symmetry. The latter two requirements then force this antilinear symmetry to be $CPT$, while forcing the Hamiltonian to be real rather than Hermitian. Our work justifies the use of the $CPT$ theorem in establishing the equality of the lifetimes of unstable particles that are charge conjugates of each other. We show that the Euclidean time path integrals of a $CPT$-symmetric theory must always be real. In the quantum-mechanical limit the key results of the $PT$ symmetry program of Bender and collaborators are recovered, with the $C$-operator of the $PT$ symmetry program being identified with the linear component of the charge conjugation operator. 
\end{abstract}
\maketitle

\section{Antilinear Symmetry and Energies}

Hermiticity of a Hamiltonian has been a cornerstone of quantum mechanics ever since its inception. Once a 
Hamiltonian is  Hermitian it follows that all of its energy eigenvalues are real and that time evolution is unitary. Moreover, with some standard additional field-theoretic assumptions one can show that in a quantum field theory a Hermitian Hamiltonian is always $CPT$ invariant. Despite this, and due primarily to the work of Bender and collaborators \cite{Bender1998,Bender2007} on non-Hermitian but $PT$ symmetric Hamiltonians ($P$ is parity, $T$ is time reversal), it has become apparent that it is possible to achieve both real eigenvalues and the time-independent evolution of Hilbert space scalar products even if a Hamiltonian is not Hermitian, provided that it instead has an antilinear symmetry such as $PT$. Consequently, while Hermiticity is sufficient to secure the reality of eigenvalues and the time-independent evolution of scalar products it is not necessary. In this paper we show that a similar situation holds for the $CPT$ theorem, with it being possible to establish invariance of a Hamiltonian under an antilinear $CPT$ transformation even if the Hamiltonian is not Hermitian. While $CPT$ symmetry is more general than $PT$ symmetry, whenever charge conjugation $C$ is separately conserved, for non-Hermitian Hamiltonians with an underlying $CPT$ symmetry one is able to recover the key results of the $PT$ symmetry program.  

The relation between eigenvalues and antilinear symmetry dates back to Wigner's work on time reversal. Specifically, if we apply some general antilinear operator $A$ to $H|\psi\rangle=E|\psi\rangle$, we obtain $AHA^{-1}A|\psi\rangle=E^*A|\psi\rangle$. Then, if $A$ commutes with $H$ we infer that either $E$ is real and $|\psi\rangle=A|\psi\rangle$, or that $E$ is complex and the eigenvectors associated with $E$ and $E^*$ transform into each other under $A$. Thus with an antilinear symmetry energies are either real or appear in complex conjugate pairs, and since nothing in this analysis requires that $H$ be Hermitian, the eigenvalues could all be real even if $H$ is not in fact Hermitian. As a case in point consider the $H=p^2+ix^3$ Hamiltonian studied in \cite{Bender1998}. While not Hermitian, this Hamiltonian does have a $PT$ symmetry (under PT $p\rightarrow p$, $x\rightarrow -x$, $i\rightarrow -i$), and it turns out (see e. g. \cite{Bender2007}) that every one of its  eigenvalues is real.

Antilinear symmetry of a Hamiltonian is more far-reaching than Hermiticity (though of course Hamiltonians can be both Hermitian and have an antilinear symmetry, as many do), and as such it provides options for quantum theory that are not allowed by Hermiticity, with antilinearity being able to encompass both decays and non-diagonalizable Jordan-block Hamiltonians such as those of relevance to fourth-order derivative theories \cite{Bender2008b}. For decays, the utility in having a complex conjugate pair of energy eigenvalues is that when a state $|A\rangle$ (the state whose energy has a negative imaginary part) decays into some other state $|B\rangle$ (the one whose energy has a positive imaginary part), as the population of state $|A\rangle$ decreases that of $|B\rangle$ increases in proportion. This interplay between the two states is found \cite{Mannheim2013} to lead to the time-independent evolution of scalar products associated with the overlap of the two states. In contrast, in theories based on Hermitian Hamiltonians, to describe  a decay one by hand adds a non-Hermitian term to the  Hamiltonian, and again by hand chooses its sign so that only the decaying mode appears. One also does this for the decays of particles that are charge conjugates of each other, and then uses the $CPT$ theorem to show that their decay rates are equal even though the standard proof of the $CPT$ theorem presupposes that the Hamiltonian is Hermitian, in which case neither of the particles would decay at all \cite{footnoteA}. In this paper we will address this issue by deriving the $CPT$ theorem without assuming Hermiticity. (Some alternate discussion of the $CPT$ theorem in the presence of unstable states may be found in \cite{Selover2013}.)

\section{Antilinear Symmetry and Time Evolution}

In the standard discussion of the time evolution generated by a time-independent Hamiltonian, one introduces states $|R_i(t)\rangle$ that evolve according to $|R_i(t)\rangle=\exp(-iHt)|R(t=0)\rangle$, with the standard Dirac scalar product $\langle R_i(t)|R_j(t)\rangle=\langle R_i(t=0)|\exp(iH^{\dagger}t)\exp(-iHt)|R_j(t=0)\rangle$ then being time independent if $H$ is Hermitian. While one can immediately conclude that the standard Dirac norm would not be time independent if $H \neq H^{\dagger}$, that does not preclude the existence of some other scalar product that would be time independent. In the more general case we note that the eigenvector equation $i\partial_{t}|R(t) \rangle=H|R(t) \rangle $ only involves the kets and serves to identify right-eigenvectors. Since the bra states are not specified by an equation that only involves the kets, there is some freedom in choosing them. As discussed for instance in \cite{Mannheim2013}, in general one should introduce left-eigenvectors of the Hamiltonian according to $-i\partial_{t} \langle L|=\langle L|H$, and use the more general norm $\langle L|R \rangle$, since for it one does have $\langle L(t)|R(t) \rangle=\langle L(t=0)|\exp(iHt)\exp(-iHt)|R(t=0) \rangle=\langle L(t=0)|R(t=0) \rangle$, so that this particular norm is preserved in time. While this norm coincides with the Dirac norm $\langle R|R \rangle$ when $H$ is Hermitian, when $H$ is not Hermitian one should use the $\langle L|R \rangle$ norm instead. 

If $|R_i(t) \rangle$ is a right-eigenvector of $H$ with some general energy eigenvalue $E_i=E_i^R+iE_i^I$, and $\langle L_j(t)|$ is a left-eigenvector of $H$ with energy eigenvalue $E_j=E_j^R+iE_j^I$, then in general we can write
\begin{eqnarray}
\langle L_j(t)|R_i(t) \rangle=\langle L_j(0)|R_i(0) \rangle e^{-i(E_i^R+iE_i^I)t+i(E_j^R-iE_j^I)t}.
\label{CPT1}
\end{eqnarray}
If these norms are to be time independent, the only allowed non-zero norms are those that obey
\begin{eqnarray}
&&E_i^R=E_j^R,\qquad E_i^I=-E_j^I,
\label{CPT2}
\end{eqnarray}
with every other norm having to obey $\langle L_j(0)|R_i(0) \rangle=0$. Thus we see that the only non-zero overlaps are precisely those associated with eigenvalues that are purely real or are in complex conjugate pairs, with this being the most general condition under which scalar products can be time independent. And with $\langle E_i^R-iE_i^I|E_i^R+iE_i^I \rangle=\exp(iE_i^Rt-E_i^It)\exp(-iE_i^Rt+E_i^It)$ being time independent, in the complex energy sector the only non-zero overlaps are precisely between a state that decays in time and one that grows in time at the complementary rate. As we had noted above, this is just as needed to maintain the time independence of the transition between them.

While we had noted above that if one has an antilinear symmetry one can establish the energy relationship given in (\ref{CPT2}), for our purposes here we need to show that if one is given (\ref{CPT2}), i. e. if one is given time-independent evolution of scalar products, then $H$ must admit of an antilinear symmetry. To this end we consider the eigenequation
\begin{eqnarray}
i\frac{\partial}{\partial t}|\psi(t)\rangle=H|\psi(t)\rangle=E|\psi(t)\rangle.
\label{CPT3}
\end{eqnarray}
On replacing the parameter $t$ by $-t$ and then multiplying by a general antilinear operator $A$ we obtain
\begin{eqnarray}
i\frac{\partial}{\partial t}A|\psi(-t)\rangle=AHA^{-1}A|\psi(-t)\rangle=E^*A|\psi(-t)\rangle.
\label{CPT4}
\end{eqnarray}
Then, because we are explicitly interested in the case where $E^*$ is an eigenvalue of $H$ we can set $HA|\psi(-t)\rangle=E^*A|\psi(-t)\rangle$, and thus obtain
\begin{eqnarray}
(AHA^{-1}-H)A|\psi(-t)\rangle=0.
\label{CPT5}
\end{eqnarray}
Now (\ref{CPT5}) has to hold for every eigenstate of $H$, and if the set of all such eigenstates is complete, we can set $[H,A]=0$ as an operator identity. We thus conclude that if all left-right scalar products are time independent, then $H$ must possess an antilinear symmetry.  To determine what that antilinear symmetry might be, with $H$ being a generator of the Poincare group,  we turn now to the implications of the complex Lorentz group.

\section{The Complex Lorentz Group}

When Lorentz transformations were first introduced into physics, they were taken to be real since one only considered transformations on real $(x,y,z,t)$ coordinates of the form $x^{\prime \mu}=\Lambda^{\mu}_{\phantom{\mu}\nu}x^{\nu}$ with real $\Lambda^{\mu}_{\phantom{\mu}\nu}$ (i. e. observer moving with a real velocity), so that the transformed coordinates would be real also. Nonetheless, if we were to take the velocity and $\Lambda^{\mu}_{\phantom{\mu}\nu}(v)$ to be complex the flat space line element $\eta_{\mu\nu}x^{\mu}x^{\nu}$ would still be invariant. 

Moreover, as well as the line element, similar remarks apply to the action $I=\int d^4x L(x)$. With $L(x)$ being a Lorentz scalar, this action is invariant under real Lorentz transformations of the form $\exp(iw^{\mu\nu}M_{\mu\nu})$ where the six $w^{\mu\nu}=-w^{\nu\mu}$ are real parameters and the six $M_{\mu\nu}=-M_{\nu\mu}$ are the generators of the Lorentz group. Specifically, with $M_{\mu\nu}$ acting on the Lorentz scalar $L(x)$ as $x_{\mu}p_{\nu}-x_{\nu}p_{\mu}$, under an infinitesimal Lorentz transformation the change in the action is  given by $\delta I=2w^{\mu\nu}\int d^4x x_{\mu}\partial_{\nu}L(x)$, and thus by $\delta I=2w^{\mu\nu}\int d^4x \partial_{\nu}[x_{\mu}L(x)]$. Since the change in the action is a total divergence, the familiar invariance of the action under real Lorentz transformations is secured. However, we now note that that nothing in this argument depended on $w^{\mu\nu}$ being real, with the change in the action still being a total divergence even if $w^{\mu\nu}$ is complex. The action $I=\int d^4x L(x)$ is thus actually invariant under complex Lorentz transformations as well and not just  under real ones, with complex Lorentz invariance thus  being a natural symmetry in physics.

Further justification for the relevance of the complex Lorentz group is provided by spinors, since they are contained not in $SO(3,1)$ itself but in its unitary and thus complex covering group. For spinor fields we can consider a ``line element" $\psi^{\rm Tr}C\psi$ (see e. g. \cite{Mannheim1985}) in Grassmann space where $\rm Tr$ denotes transpose in the Dirac gamma matrix space and $C$ is the Dirac gamma matrix that effects $C^{-1}\gamma^{\mu}C=-\gamma^{\mu}_{\rm Tr}$. With a Dirac spinor transforming as $\psi \rightarrow \exp(iw^{\mu\nu}M_{\mu\nu})\psi$, we see that since $\psi^{\rm Tr}C\psi$ does not involve Hermitian conjugation, it is invariant not just under real but also complex $w^{\mu\nu}$. Now as it stands the scalar quantity $\bar{\psi}\psi=\psi^{\dagger}\gamma^0\psi$ would  be invariant under real Lorentz transformations but not under complex ones. However, as will be central to our discussion below of charge conjugation, we note that since a Dirac spinor is reducible under the Lorentz group we can decompose it as $\psi=\psi_{\rm A}+i\psi_{\rm B}$, where $\psi_{\rm A}$ and $\psi_{\rm B}$  are self-conjugate Majorana spinors that in the Majorana representation of the Dirac gamma matrices (see e. g. \cite{Mannheim1984}) obey $\psi_{\rm A}^{\dagger}=\psi_{\rm A}$ and $\psi_{\rm B}^{\dagger}=\psi_{\rm B}$. With this decomposition we understand a complex Lorentz transformation to be implemented on the separate $\psi_{\rm A}$ and $\psi_{\rm B}$, with $\bar{\psi}\psi$ then being invariant under complex Lorentz transformations too. Thus in the following we shall consider the implications of complex Lorentz invariance.

Complex Lorentz invariance is of significance to both $PT$ and $CPT$ transformations, and both will be needed for the $CPT$ theorem, since under $CPT$ the argument of a field changes from $x^{\mu}$ to $-x^{\mu}$, just as required by  the $PT$ part of the $CPT$ transformation. For $PT$ transformations first, we note that on applying the specific sequence of Lorentz boosts: first $x^{\prime}=x\cosh\xi +t\sinh \xi $, $t^{\prime}=t\cosh\xi +x\sinh \xi $, then $y^{\prime}=y\cosh \xi +t\sinh \xi $, $t^{\prime}=t\cosh\xi +y\sinh \xi $, and finally $z^{\prime}=z\cosh \xi +t\sinh \xi $, $t^{\prime}=t\cosh\xi +z\sinh \xi $, each with a complex boost angle $\xi=i\pi$, we generate $(x,y,z,t)\rightarrow (-x,-y,-z,-t)$. On defining  $\pi\tau=\Lambda^{0}_{\phantom{0}3}(i\pi)\Lambda^{0}_{\phantom{0}2}(i\pi)\Lambda^{0}_{\phantom{0}1}(i\pi)$,  $\pi\tau$ effects $\pi\tau:x^{\mu}\rightarrow -x^{\mu}$. However, though this transformation does  indeed reverse the signs of all four of the coordinates just as a $PT$ transformation does, $\pi\tau$ itself is not the $PT$ transformation of interest to physics since time reversal has to be an antilinear operator rather than a linear one. Nonetheless, we can always represent an antilinear operator as a linear operator times complex conjugation. On introducing an operator  $K_T$ that conjugates complex numbers, up to intrinsic system-dependent phases we can then set $PT=\pi\tau K_T$, i. e. we can represent $PT$ as a complex Lorentz boost times complex conjugation, to thus give a $PT$ transformation an association with the complex Lorentz group \cite{footnoteB}.

With  $C$, $P$, and $T$ respectively acting on spinors as $1$, $\gamma^0$, and $\gamma^1\gamma^2\gamma^3$ in the Majorana basis of the Dirac gamma matrices, for spinors  $CPT$ effects  $CPT\psi(x)[CPT]^{-1}=-i\gamma^5\psi^{\dagger}(-x)$. Then with $M^{0i}=i[\gamma^0,\gamma^i]/4$, $\Lambda^{0i}(i\pi)=\exp(-i\pi\gamma^0\gamma^i/2)=-i\gamma^0\gamma^i$, quite remarkably we find that in the Dirac gamma matrix space we recognize the previously introduced complex Lorentz transformation $\Lambda^{0}_{\phantom{0}3}(i\pi)\Lambda^{0}_{\phantom{0}2}(i\pi)\Lambda^{0}_{\phantom{0}1}(i\pi) =i\gamma^0\gamma^1\gamma^2\gamma^3=\gamma^5$ as being none other than the linear part of a $CPT$ transformation in spinor space, with $CPT$ thus having a natural connection to the complex Lorentz group. 

As well as identify the linear part of a $CPT$ transformation we also need to consider its conjugation aspects, and initially it would appear that $C$ would differ from $T$ since $T$ involves complex conjugation of complex numbers while $C$ involves charge conjugation of quantum fields. However, the two types of conjugation can actually be related, since charge conjugation converts a field into its Hermitian conjugate, and Hermitian conjugation does conjugate factors of $i$. If for instance we consider a charged scalar field $\phi(x)$, then under $C$ it transforms as $\phi(x)\rightarrow C\phi(x)C^{-1}=\phi^{\dagger}(x)$. However, suppose we break $\phi(x)$ into two Hermitian components according to $\phi(x)=\phi_1(x)+i\phi_2(x)$. Now since $C$ effects $\phi(x)\rightarrow \phi^{\dagger}(x)$, we can achieve this in two distinct ways. We can have $C$ act linearly on $\phi_1(x)$ and $\phi_2(x)$ according to $\phi_1(x)\rightarrow \phi_1(x)$, $\phi_2(x)\rightarrow -\phi_2(x)$ while having no effect on the factor of $i$, or we can have $C$ act antilinearly on $i$ according to  $i\rightarrow CiC^{-1}=-i$ while having no effect on the Hermitian $\phi_1(x)$ and $\phi_2(x)$. For our purposes here the latter interpretation is not only the more useful as it helps us keep track of factors of $i$ in quantities such as $\phi_{\pm}(x)=\phi_1(x)\pm i\phi_2(x)$, as we will see below, it will prove crucial to our derivation of the $CPT$ theorem.  Moreover, we note that with an antilinear interpretation for $C$ we do not even need $\phi_1(x)$ and $\phi_2(x)$ to actually be Hermitian fields at all. They could instead, for instance,  be defined as being self-conjugate under $C$ or self-conjugate under $CPT$.

In addition to the complex conjugation effected by $K_C$, $C$ could also effect a linear transformation $\kappa$ as well, and so we can write $CPT$ as  $\kappa\pi\tau K$, where $K=K_CK_T$ complex conjugates everything it acts on, c-numbers and q-numbers alike \cite{footnoteD}.  With this analysis also holding for  Majorana spinors (cf. $\psi=\psi_{\rm A}+i\psi_{\rm B}$), and with Majorana spinors being able to serve as the fundamental representation of the Lorentz group (a Majorana spinor can be written as a Weyl spinor plus its charge conjugate \cite{Mannheim1984}), we can represent $CPT$ as the generic $\kappa\pi\tau K$ when acting on any representation of the Lorentz group.

While $C$ as defined here  effects $C\phi_{\pm}(x)C^{-1}=\phi_{\mp}(x)$, $C$ does not complex conjugate the individual $\phi_i(x)$ themselves. However,  $T$ still can, and in fact must, since the $[x,p]=i$  commutator for instance is preserved under $T$ according to $x\rightarrow x$, $p\rightarrow -p$, $i \rightarrow -i$. To see how $T$ explicitly achieves this, we set $x=(a+a^{\dagger})/2^{1/2}$, $p=i(a^{\dagger}-a)/2^{1/2}$, $[a,a^{\dagger}]=1$. Thus we need $T$ to effect $a\rightarrow a$, $a^{\dagger}\rightarrow a^{\dagger}$, $i \rightarrow -i$, and this is achieved by the antilinear $K_T$. In the Fock space labelled by $|\Omega\rangle$, $a^{\dagger}|\Omega\rangle$, ..., where $a|\Omega\rangle=0$, $a$ and $a^{\dagger}$ can both be represented by  infinite-dimensional matrices that are purely real. With $x$ being real and symmetric and $p$ being pure imaginary and anti-symmetric in this Fock space, in this Fock space only the $i$ in the operator $p$ is affected by $T$. With the same analysis also holding for commutators of the generic form $[\phi(\bar{x},t=0),\pi(\bar{x}^{\prime},t=0)]=i\delta^3(\bar{x}-\bar{x}^{\prime})$, we see that due to our treating $C$ as antilinear, for every function $F$ that is built out of canonical quantum fields,  it follows that $KFK^{-1}=F^*$ (i. e. $K_CK_T$ conjugates all factors of $i$). It is this specific feature that will enable us to derive the $CPT$ theorem.

\section{Derivation of the $CPT$ Theorem}

As noted for instance in \cite{Weinberg1995}, under $CPT$ every  irreducible representation of the Lorentz group transforms as $CPT\phi(x)[CPT]^{-1}= \eta(\phi) \phi^{\dagger}(-x)$ with a $\phi$-dependent phase $\eta(\phi)$ that depends on the spin of each $\phi$ and obeys $\eta^2(\phi)=1$; with spin zero fields (both scalar and pseudoscalar) expressly having $\eta(\phi)=1$ \cite{footnoteE}. Since the most general Lorentz invariant Lagrangian must be built out of sums of appropriately contracted spin zero products of fields with arbitrary numerical coefficients, and since it is only spin zero fields that can multiply any given net spin zero product an arbitrary number of times and still yield net spin zero, all net spin zero products of fields must have a net $\eta(\phi)$ equal to one \cite{footnoteF}. Generically, such products could involve $\phi\phi$ or $\phi^{\dagger}\phi$ type contractions. However, requiring the Lagrangian and thus the Hamiltonian to be Hermitian then forces the contractions to be Hermitian (only $\phi^{\dagger}\phi$) while forcing the  coefficients to all be real, with the Hamiltonian then being  $CPT$ invariant \cite{Weinberg1995}. 

In order to extend the $CPT$ theorem to non-Hermitian Hamiltonians, we note first that even in the  non-Hermitian case Lorentz invariance still requires every term in the Lagrangian to have a net $\eta(\phi)$ equal to one. With $CPT$ effecting $CPT\phi_{\pm}(x)[CPT]^{-1}= \eta(\phi)\phi_{\mp}(-x)$ we will need some non-Hermitian-based  reason in order to be able to exclude any $\phi_{\pm}(x)\phi_{\pm}(x)$ type contractions. To this end we note that with the linear part of a $CPT$ transformation having been identified as the particular complex Lorentz transformation $\Lambda^{0}_{\phantom{0}3}(i\pi)\Lambda^{0}_{\phantom{0}2}(i\pi)\Lambda^{0}_{\phantom{0}1}(i\pi)$, under this transformation every net spin zero term in a Lorentz invariant action $I=\int d^4x{{\cal L}}(x)$ will transform so that $I\rightarrow \int d^4x{{\cal L}}(-x)=\int d^4x{{\cal L}}(x)=I$, with the action thus being left invariant. Then, under the full $CPT$ transformation, and precisely because of our having taken $C$ to be antilinear, every term in the action will transform so that $I \rightarrow \int d^4x K{{\cal L}}(x)K^{-1}$. Thus under a $CPT$ transformation, the full Hamiltonian will transform as $H \rightarrow KHK^{-1}$. Since at this point we have now arrived at (\ref{CPT4}) with $A$ being identified with $K$, we see that the requirement of time-independent evolution of scalar products given in (\ref{CPT5}) then follows, with $H$ obeying $H =KHK^{-1}$. The $CPT$ invariance of $H$ is thus secured, with there thus being only $\phi_{\pm}(x)\phi_{\mp}(x)$ type terms and no $\phi_{\pm}(x)\phi_{\pm}(x)$ type ones allowed in $H$, and with all coefficients again being real  \cite{footnoteG}. With our definition of $K$ we see that $H$ obeys $H=H^*$, while not being required to obey $H=H^{\dagger}$. The $CPT$ theorem is thus extended to non-Hermitian Hamiltonians that generate time-independent evolution of scalar products \cite{footnoteH}.

\section{Implications}

When a time-independent Hamiltonian is real (as would be the case if $H=KHK^{-1}=H^*$), for Euclidean times $\tau=it$ the time evolution operator $\exp(-iHt)=\exp(-H\tau)$ is real. Consequently, the associated Euclidean time path integrals and Green's functions are real too.  Even though the Euclidean time path integral is real that does not mean that all energy eigenvalues are necessarily real, since if they appear in complex conjugate pairs and have complex conjugate wave functions, the Euclidean time path integral would still be real. In fact this is the most general way in which the Euclidean path integral could be real if the Hamiltonian is not Hermitian, and is just as required of antilinear $CPT$ symmetry.

We had earlier referred to the $PT$ symmetric $H=p^2+ix^3$. Since it is the quantum-mechanical  limit of a relativistic theory with appropriate Hamiltonian density $H=\Pi^2+i\Phi^3$, it is $CPT$ invariant. With $\Phi$ being uncharged, this $H$ is separately $\kappa K_C$ invariant, and thus it is indeed $PT$ symmetric. Now we can realize the $[x,p]=i$ commutator by $x=i(b-b^{\dagger})/2^{1/2}$, $p=(b^{\dagger}+b)/2^{1/2}$ where $[b,b^{\dagger}]=1$. (This realization is unitarily equivalent to  $x=(a+a^{\dagger})/2^{1/2}$, $p=i(a^{\dagger}-a)/2^{1/2}$.) In the Fock space where $b|\Omega\rangle =0$, $x$ is pure imaginary and antisymmetric, $p$ is real and symmetric, and thus even though $H$ is not Hermitian, in this particular occupation number space $H=p^2+ix^3$ can be represented by an infinite-dimensional matrix all of whose elements are real. Hence, despite its appearance $H=p^2+ix^3$ obeys $H=H^*$ \cite{footnoteI}.

As an example of a Hamiltonian that is $CPT$ invariant while having complex conjugate energy pairs, consider the fourth-order Pais-Uhlenbeck two-oscillator ($p_z,z$ and  $p_x,x$) model studied in \cite{Bender2008a,Bender2008b}. Its Hamiltonian is given by $H_{\rm PU}=p_x^2/2\gamma+p_zx+\gamma\left(\omega_1^2+\omega_2^2 \right)x^2/2-\gamma\omega_1^2\omega_2^2z^2/2$ where initially $\omega_1$ and $\omega_2$ are real (this Hamiltonian is the quantum-mechanical limit of a covariant fourth-order neutral scalar field theory \cite{Bender2008b}). $H_{\rm PU}$ turns out not to be Hermitian but to instead be $PT$ symmetric \cite{Bender2008a,Bender2008b}, with all energy eigenvalues nonetheless being given by the real $E(n_1,n_2)=(n_1+1/2)\omega_1+(n_2+1/2)\omega_2$. In addition, $H_{\rm PU}$ is $CPT$ symmetric since $C$ plays no role ($[\kappa K_C,H_{PU}]=0$), while thus descending from a neutral scalar field theory that is also $CPT$ invariant. If we now set  $\omega_1=\alpha+i\beta$, $\omega_2=\alpha-i\beta$ with real $\alpha$ and $\beta$, we see that $(\omega_1^2+\omega_2^2)/2=\alpha^2-\beta^2$ and  $\omega_1^2\omega_2^2=(\alpha^2+\beta^2)^2$ both remain real. In consequence $H_{\rm PU}$ remains $CPT$ invariant, but now the energies come in complex conjugate pairs as per $E(n_1,n_2)=(n_1+1/2)(\alpha+i\beta)+(n_2+1/2)(\alpha-i\beta)$. It is also of interest to note that when $\omega_1=\omega_2=\alpha$  with $\alpha$ real, the Hamiltonian becomes of non-diagonalizable, and thus of manifestly non-Hermitian, Jordan-block form \cite{Bender2008b}, with its $CPT$ symmetry not being impaired. Thus for $\omega_1$ and $\omega_2$ both real and unequal, both real and equal, or being complex conjugates of each other, in all cases one has a non-Hermitian but $CPT$-invariant Hamiltonian that descends from a quantum field theory whose Hamiltonian while not Hermitian is nonetheless $CPT$ symmetric.

The $PT$ studies of Bender and collaborators are mainly quantum-mechanical ones in which the field-theoretic charge conjugation operator plays no role. In these studies it has been found \cite{Bender2007} that as well as be $PT$ symmetric, the Hamiltonian is also symmetric under a specific discrete linear operator also called $C$, which obeys $[C,H]=0$ and  $C^2=1$. In addition, this $C$ obeys $[C,PT]=0$ when all energies are real, and obeys  $[C,PT]\neq 0$ when energies are in complex pairs \cite{Bender2010}. The $CPT$ symmetry of any given relativistic theory ensures the $PT$ symmetry of any $C$-invariant quantum-mechanical theory that descends from it, while guaranteeing that it must possess a linear operator, viz. our previously introduced $\kappa$, that obeys $[\kappa,H]=0$ and $\kappa^2=1$, and so we can now identify $\kappa$ (or a similarity transform of it) with the $C$ operator of $PT$ theory \cite{footnoteJ}. Our work thus puts the $PT$ symmetry studies of theories with non-Hermitian Hamiltonians on a quite secure quantum-field-theoretic foundation. 

\section{Applications}

Once one extends the $CPT$ theorem to non-Hermitian Hamiltonians, the most interesting applications of our ideas are to  situations that can never be encompassed by Hermitian Hamiltonians. The currently most explored such area is in applications of $PT$ symmetry in the complex conjugate energy pair situation, where there are both growing and decaying modes. In the $PT$ literature such modes are referred to as gain and loss, with many experimental examples having been identified \cite{Special2012,Theme2013}.  Moreover, typically in these cases, as one adjusts parameters one can transition to the region where all eigenvalues are real. At the point of the transition, known as an exceptional point in the $PT$ literature, the Hamiltonian becomes of a  non-diagonalizable, and thus manifestly non-Hermitian, Jordan-block form, and experimental effects due to exceptional points have also been discussed in the literature. 

For relativistic quantum theory our results can be applied to particle decays such as those encountered in the $K$ meson system.  Specifically, the time-independent transition matrix elements that we obtain precisely provide for probability conserving transitions between decaying states and the growing states into which they decay, with the $CPT$ theorem that we have derived here then requiring that the transition rates for the decays of particles and their antiparticles be equal. To be more specific, we note that as well as provide an explicitly solvable model that is non-Hermitian but $CPT$ invariant, the two-oscillator Pais -Uhlenbeck model can also serve as a prototype for discussing decays. In the region where $\omega_1=\alpha+i\beta$,  $\omega_2=\alpha-i\beta$, the Hamiltonian is given by the $CPT$-symmetric $H_{\rm PU}=p_x^2/2\gamma+p_zx+\gamma\left(\alpha^2-\beta^2 \right)x^2-\gamma(\alpha^2+\beta^2)^2z^2/2$. Not only does this model contain both decaying ($\omega_2=\alpha-i\beta$) and growing modes ($\omega_1=\alpha+i\beta$), as per (\ref{CPT1}) and (\ref{CPT2}) it leads to time-independent transitions between them, and thus describes the decay of one mode into the other. If we now make the momentum and position operators be complex, which we can do in a charge conjugation invariant manner, the field-theoretic generalization of the model will then contain both particles and antiparticles, with the $CPT$ invariance of the Hamiltonian then requiring that the decay rates for particles and their antiparticles be equal.

Another case that cannot be described by a Hermitian Hamiltonian is encountered in the currently viable fourth-order derivative conformal gravity theory, a conformal invariant, general coordinate invariant theory of gravity that has been advanced as candidate alternate theory of gravity \cite{Mannheim2012}. The conformal gravity action is given by $I_{\rm W}=-\alpha_g\int d^4x (-g)^{1/2}C_{\lambda\mu\nu\kappa} C^{\lambda\mu\nu\kappa}$ where $C^{\lambda\mu\nu\kappa}$ is the Weyl conformal tensor, and its Hamiltonian is a relativistic generalization of the equal-frequency Pais-Uhlenbeck Hamiltonian \cite{Mannheim2011a}. Consequently, the conformal gravity Hamiltonian is  of a non-Hermitian, non-diagonalizable, Jordan-Block form \cite{Bender2008b,Mannheim2011a}, to thus serve as an explicit field-theoretic example of a Hamiltonian that is not Hermitian but is  $CPT$ symmetric.

\begin{acknowledgments}
The author wishes to thank  Dr. Carl Bender for some very helpful discussions.
\end{acknowledgments}


\begin{thebibliography}{99}
\bibitem{footnoteA} Given a Hermitian weak interaction Hamiltonian $H_{\rm W}$, through standard use of the $CPT$ theorem one can show the equality of matrix elements such as $\langle K_0|H_{\rm W}|\pi^+\pi^-\rangle$ and $\langle \bar{K}_0|H_{\rm W}|\pi^+\pi^-\rangle$. With the Hamiltonian being Hermitian, neither kaon should decay. And so to force them to do so one chooses a boundary condition (decaying mode only and no growing mode) that is not $CPT$ invariant.
\bibitem{footnoteB} We note that no such complex association exists for either $P$ ($\vec{x}\rightarrow -\vec{x}$) or $T$ ($t \rightarrow -t$) separately, since, with each Lorentz transformation involving an even number of spacetime coordinates, there is no transformation that could reverse the sign of an odd number of them.
\bibitem{footnoteE} Since $[\phi(\bar{x},t=0),\pi(\bar{x}^{\prime},t=0)]=i\delta^3(\bar{x}-\bar{x}^{\prime})$,  
the conjugate $\pi(x)$ always has the opposite $CPT$ to $\phi(x)$.
\bibitem{footnoteF} With a fermion in the fundamental representation of the Lorentz group transforming as $CPT\psi(x)[CPT]^{-1}= -i\gamma^5 \psi^{\dagger}(-x)$, one can anticipate that every irreducible representation of the Lorentz  group will obey an analogous relation since $\gamma^5$ commutes with the Lorentz generators $i[\gamma^{\mu},\gamma^{\nu}]/4$. With the $CPT$ phases of $\bar{\psi}\psi$, $\bar{\psi}i\gamma^5\psi$, $\bar{\psi}\gamma^{\mu}\psi$, $\bar{\psi}\gamma^{\mu}\gamma^5\psi$, and $\bar{\psi}i[\gamma^{\mu},\gamma^{\nu}]\psi$ respectively being $+1$, $+1$, $-1$, $-1$, and $+1$, for integer spin representations of the Lorentz group $CPT$ phases alternate with spin, so that spin zero combinations (combinations that must always contain even numbers of fermions) will all have a $CPT$ phase equal to $+1$. No analogous statement can be made for any other transformation involving $C$, $P$, or $T$.
\bibitem{footnoteG} We note that even with antilinear operators, commutation relations are preserved under similarity transforms. If we set $H^{\prime}=SHS^{-1}$, $K^{\prime}=SKS^{-1}=S[S^{-1}]^*K=KS^*S^{-1}$, $K^{\prime 2}=I$,  then $[K^{\prime},H^{\prime}]=S[K,H]S^{-1}$, with the operator $K$ thus transforming along with $H$. In consequence, the commutation relation $[CPT,H]=0$ is preserved under a similarity transform.
\bibitem{footnoteI} Starting from an initial $CPT$ (and $PT$) symmetric and thus real Hamiltonian density $H=-\Pi^2+\Phi^3$ with a $CPT$ even neutral pseudoscalar field $\Phi$ and its $CPT$ odd conjugate $\Pi$, the substitution $\Phi\rightarrow -i\Phi\equiv -ix$, $\Pi\rightarrow i\Pi\equiv ip$ yelds the $CPT$ (and $PT$) odd $x$, and the $CPT$ (and $PT$) even $p$, and the still $CPT$ (and $PT$) symmetric and still real $H\rightarrow \Pi^2+i\Phi^3 \equiv p^2+ix^3$.
\bibitem{footnoteJ} Our point here is that we can use joint $CPT$ and $C$ invariance to obtain $PT$ invariance even if $H$ is non-Hermitian.
\end{thebibliography}
\end{document}